\documentclass{elsart}

\usepackage{graphicx}
\usepackage{amssymb}

\journal{Physics Letters A}

\begin{document}

\begin{frontmatter}

% Title, authors and addresses

% use the thanksref command within \title, \author or \address for footnotes;
% use the corauthref command within \author for corresponding author footnotes;
% use the ead command for the email address,
% and the form \ead[url] for the home page:
% \title{Title\thanksref{label1}}
% \thanks[label1]{}
% \author{Name\corauthref{cor1}\thanksref{label2}}
% \ead{email address}
% \ead[url]{home page}
% \thanks[label2]{}
% \corauth[cor1]{}
% \address{Address\thanksref{label3}}
% \thanks[label3]{}

\title{Force measurements of a superconducting-film actuator for a cryogenic
       interferometric gravitational-wave detector}

% use optional labels to link authors explicitly to addresses:
% \author[label1,label2]{}
% \address[label1]{}
% \address[label2]{}

\author[KEK]{N. Sato\corauthref{cor1}},
\corauth[cor1]{Corresponding author.}
\ead{saton@post.kek.jp}
\author[KEK]{T. Haruyama},
\author[MUE]{N. Kanda\thanksref{now3}},
\thanks[now3]{Present address: Osaka City University, 3-3-138 Sugimoto,
              Sumiyoshi, Osaka 558-8585, Japan}
\author[ICRR]{K. Kuroda},
\author[ICRR]{S. Miyoki},
\author[ICRR]{M. Ohashi},
\author[KEK]{Y. Saito},
\author[KEK]{T. Shintomi},
\author[KEK]{T. Suzuki},
\author[ICRR]{D. Tatsumi\thanksref{now1}}
\thanks[now1]{Present address: National Astronomical Observatory, 2-21-1 Osawa,
              Mitaka, Tokyo 181-8588, Japan}
\author[ICRR]{C. Taylor\thanksref{now2}},
\thanks[now2]{Present address: Rio Tinto Australian Science Olympiads Box 7251,
             Canberra Mail Centre ACT 2610, Australia}
\author[KEK]{T. Tomaru},
\author[KEK]{T. Uchiyama},
\author[KEK]{A. Yamamoto}

\address[KEK]{High Energy Accelerator Research Organization (KEK),
              1-1 Oho, Tsukuba, Ibaraki 305-0801, Japan}
\address[MUE]{Miyagi University of Education, Aramaki Aza Aoba, Aoba, Sendai,
              Miyagi 980-0845, Japan}
\address[ICRR]{ICRR, The University of Tokyo, 5-1-5 Kashiwanoha, Kashiwa,
               Chiba 277-8582, Japan}

\begin{abstract}

	We measured forces applied by an actuator with a YBa$_2$Cu$_3$O$_7$ 
(YBCO) film at near 77 K for the Large-scale Cryogenic Gravitational-wave
Telescope (LCGT) project.
 An actuator consisting of both a YBCO film of 1.6 $\mu$m thickness 
and 0.81 cm$^2$ area and a solenoid coil exerted a force of up to 0.2 mN 
on a test mass.
 The presented actuator system can be used to displace the mirror 
of LCGT for fringe lock of the interferometer.

\end{abstract}

\end{frontmatter}

% main text
%%% \section{}
%%% \label{}

\section{Introduction}

	Conventional actuators for fringe lock of interferometric 
gravitational-wave detectors consist of solenoid coils and 
permanent magnets glued to test masses.
 Previous studies showed that the mechanical quality factor ($Q$) is always 
degraded considerably whenever any materials are attached to the test mass
\cite{Gillespie,Yamamoto}.
 This is not a serious problem for current interferometric 
gravitational-wave detectors.
 On the other hand, a future project of the underground km-scale LCGT 
\cite{Kuroda} needs the best possible $Q$ value to reduce the thermal noise 
with cryogenic
mirrors on the test masses of sapphire substrates, because the amplitude
of the thermal noise is proportional
to $ \sqrt{T/Q} $, where $T$ is the temperature of the mirror.
 That noise reduction enables LCGT to improve the sensitivity
\cite{Uchiyama1,Uchiyama2,Uchiyama3} by one order better than 
the typical km-scale ground-based detectors.
 The attainable sensitivity depends on the amount of 
materials of the actuator parts attached to the sapphire substrate 
of the mirror.

	In this paper we present a cryogenic actuator with
a superconducting thin film of 1.6 $\mu$m thickness attached 
to a test mass driven by a solenoid coil.
 We mainly studied forces applied to the test mass by this cryogenic actuator.

\section{Experiment}
  \subsection{Estimation of force}
\begin{figure}
\includegraphics[width=\linewidth]{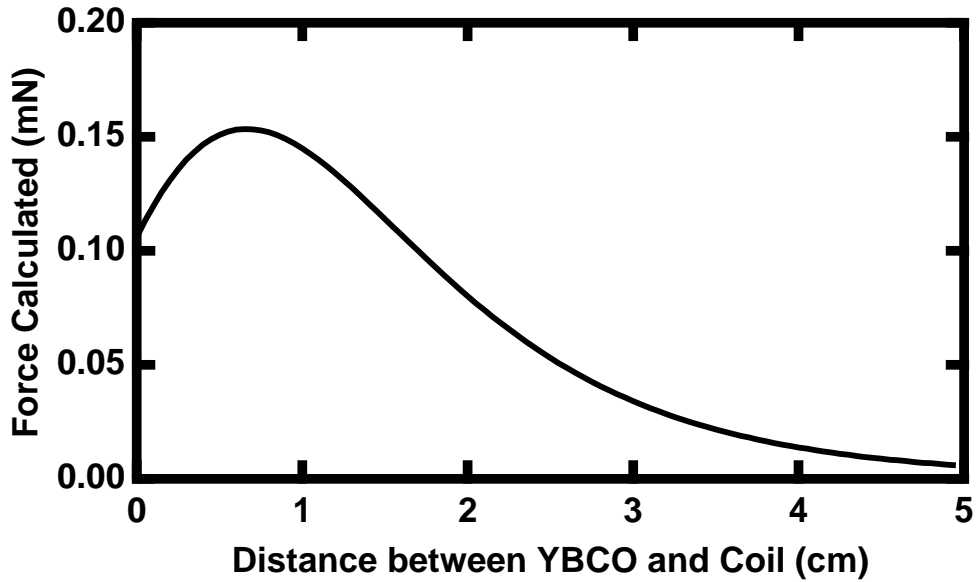}
	\caption{Example of calculated forces by a superconducting film actuator
	         as a function of the distance between the film and the nearest end
	         of the solenoid coil.
	         The solenoid coil is the same as that used by an experiment
	          with a current of 500 mA.}
	\label{fig:calforce}
\end{figure}

	When a superconducting film disk of diameter $D$
is set perpendicular to an applied magnetic field 
($B_a$) along the symmetric axis ($x$) of a solenoid coil, its magnetic moment
is obtained analytically \cite{Landau}:
$-(D^3/3) B_a$, with the assumptions
of perfect diamagnetism of the film and a weak $B_a$ so that the magnetic 
moment does not reach the saturation value at the critical current density
of the film.
 The repulsive force ($F_{sc}$) in the direction of $x$
is estimated to be

\begin{equation}
 F_{sc} \thickapprox \frac {1}{\mu _0} \left( - \frac{D^3}{3} 
  B_a \right) \frac { \partial B_a }{ \partial x }, 
\label{eq:calforce}
\end{equation}
where $\mu_0$ is the magnetic permeability of free space.
The distance dependence of the forces calculated with Eq.(\ref{eq:calforce})
is shown in Fig.~\ref{fig:calforce}.
 The positions of the superconducting film and the solenoid coil 
were arranged so that the force was applied to nearly  the maximum value
according to Fig.~\ref{fig:calforce}.

  \subsection{Method}
  
In order to measure the forces of a cryogenic actuator, 
we used a pendulum motion
with a superconducting film glued to a test mass, which was 
driven by a solenoid coil with a DC current. When a current
passes through the solenoid coil, the test mass is pushed
by $\Delta x$ and the time dependence of the position of the test mass is
changed from a sinusoidal wave with a frequency of about 1Hz 
to a sinusoidal wave plus $\Delta x$.
The displacement ($\Delta x$) is related to $ F_{sc} $ 
through the equation of
pendulum motion on the assumption of constant $ F_{sc} $,  as follows:

\begin{equation}
\Delta x = \frac {\ell F_{sc} }{m g}, \label{eq:dx}
\end{equation}
where $ \ell $ is the pendulum length, $m$ is the mass of
the test mass, and $g$ is the gravitational acceleration constant.
 We estimate $ F_{sc} $ from Eq.(\ref{eq:dx}) for the measured $ \Delta x $.
	\subsection{Experimental setup}
  
\begin{figure}
\includegraphics[width=\linewidth]{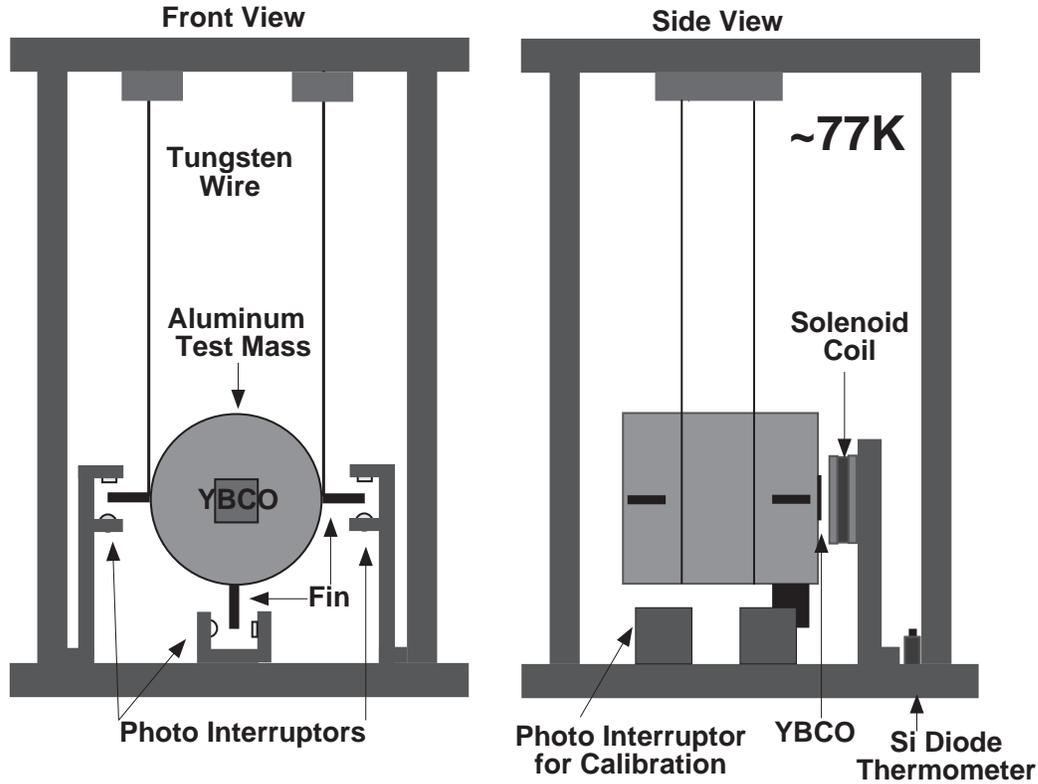}
	\caption{Schematic view of the apparatus in the vacuum chamber
	         for measuring the forces by the superconducting film actuator.
	         The pendulum, solenoid coil, and photo-interruptors
	         were fixed on a copper frame, which was thermally connected
	         to liquid nitrogen. }
	\label{fig:setup}
\end{figure}

Figure ~\ref{fig:setup}~ shows the experimental setup for a force measurement
of the superconducting film actuator. 
This setup simulated a part of a 
pendulum of the Fabry-Perot cavity of an interferometric 
gravitational-wave detector.
    A cylindrical aluminum test mass had dimensions of 5 cm in diameter
and 6 cm in length. It was suspended by two tungsten wires.
The length of the pendulum was about 20 cm.
    Superconducting film of 0.9 cm$\times$0.9 cm area was glued to one
end of the center of the test mass by epoxy glue (Stycast2850FT).
	The film was a single-sided deposition of superconducting material
on a substrate, and the side glued was the deposited one. 

	A solenoid coil wound with a wire of 0.4 mm in diameter
was 2.2 cm in inner diameter, 4.4 cm in outer diameter, and 1 cm in length. 

	On the barrel of the test mass, five rectangular fins were attached,
four of which were for monitoring the position and alignment in relation to the
solenoid coil,
and one of which was for measuring the displacement of the test mass
along the direction of the force.

A fin was set between an infrared LED and an InGaAs photodiode with an 
effective area diameter
of 1 mm. They made a photo-interruptor.
The fin interrupted a part of light emitted by the LED 
according to the motion of the test mass, and the photodiode detected
light whose intensity was proportional to the displacement of the fin.    
  Another LED-photodiode system was attached to the bottom plate. 
It was used for calibrating the temperature dependence of the light intensity 
of the LED. 
This photo-interruptor monitored the light intensity corresponding to that 
when a fin was completely outside the detection area of the photodiode.

	The pendulum, position sensors, and solenoid coil were installed 
on a copper frame in a vacuum chamber and cooled to liquid-nitrogen 
temperature in a cryostat.

A constant current of 30 mA was provided to the LEDs 
by constant-current drivers.
The output currents of photodiodes were converted to voltages by operational
amplifiers. They were fed to low-pass filters of 0.1 Hz cutoff frequency 
to extract the DC component of $ \Delta x $ from the time varying 
displacement of the test mass.
   This DC component output was read  with a digital multimeter.
All of the converted voltages of the photodiodes' outputs were
monitored with a chart recorder.
  
  \subsection{Estimation of $\Delta x$ from the photodiode output}
  
  The relation between the output of a photodiode in a photo-interruptor and
the position of a fin was measured in air at room temperature without a coil
current, and was fitted as a linear function.
The slope ($ C_{fit} $) of a fitted straight line depended on
the photo-interruptor.
	For five photo-interruptors the slopes ranged from
0.52 $ \mu m $/mV to 0.65 $ \mu m $/mV.  The fitted range of the position was 
400 $ \mu m $, in which linearity was confirmed by a simple geometrical
calculation. 
	Even if the fitting range had been changed, the difference in the slope 
would have been a few \%.

	For the coil current ($I$) at the cryogenic temperature ($T$), 
the output voltage ($V(I,T)$) from a photodiode was converted
to the corresponding output at room temperature, 
assuming that (1) the temperature was
the same for all photo-interruptors, and (2) the temperature dependence of
the photo-interruptor output was the same. 
Thus, with the output $V_{clb}(I,T)$  of the calibration photo-interruptor at
the cryogenic temperature and $V_{clb}(0, 300K)$
at room temperature,
the displacement of the test mass was estimated as follows:  

\begin{equation}
(\Delta x)_{est} = C_{fit} \left( \frac { V_{clb}(0, 300K)}
{ V_{clb}(I,T)} V(I,T) - \frac { V_{clb}(0, 300K)}
{V_{clb}(0,77K)} V(0, 77K) \right).
 \label {eq:calib}
\end{equation}
In Eq.(\ref{eq:calib}), $T$ was not always 77 K, 
but depended on the coil current. 

  \subsection{Selection of a superconducting film}

The temperature of sapphire substrates for the LCGT mirrors cannot be
decreased below 20 K when they are linked thermally 
to liquid-helium temperature
through a heat link system and a sapphire fiber suspension,
since the heat, which is generated by partial absorption of the laser beam
in the substrates,  transfers through only conduction of the
fibers \cite{Uchiyama1,Kuroda}.
	Then, the critical temperature of the superconducting film is required to 
be greater than 20 K, because it is attached to the substrate.
	Thus, we selected a commercially available high-temperature superconducting
thin film of YBCO.
The YBCO film which we used for the experiment was deposited
by the reactive thermal co-evaporation technique
on a MgO substrate, and was 1.6 $ \mu m $ thick, 0.9 cm$ \times $0.9 cm 
in area, 
88.6 K in critical temperature, and 2.7 MA/cm$^2$ in critical current density 
\cite{THEVA}.

\section{Results}

\begin{figure}
\includegraphics[width=\linewidth]{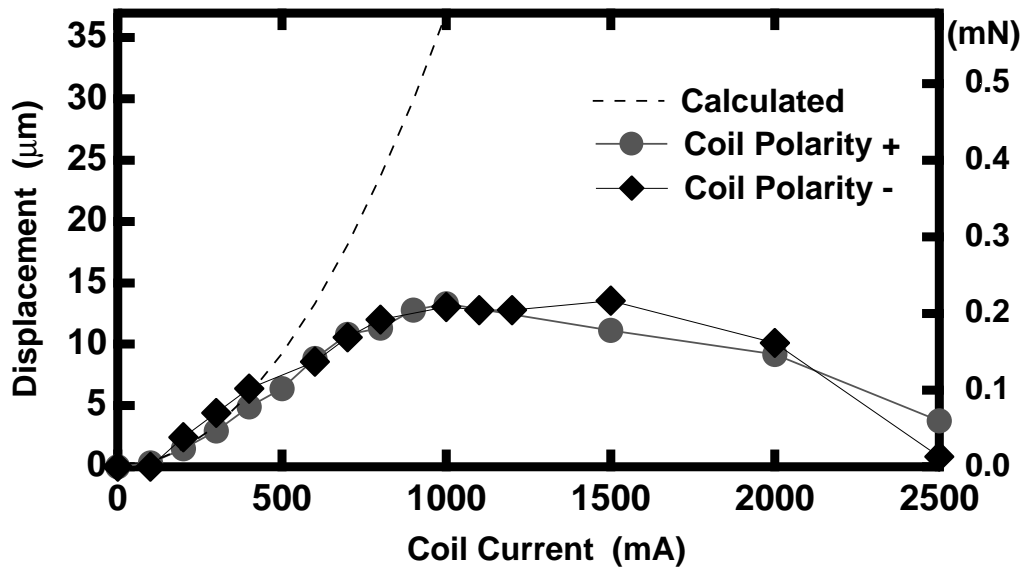}
\caption{Displacement of a test mass pushed by the YBCO film actuator 
	         as a function of the coil current. The numerical values on
	         the right-hand vertical axis indicate the forces 
	         corresponding to the displacement.}
\label{fig:force}
\end{figure}

	Figure \ref{fig:force} shows the dependence of the displacement of a test
mass on the current flowing in the solenoid coil.
At the position of the YBCO film 
the magnetic field was about 3.5 mT for a 500 mA coil current. 
Using Eq.(\ref{eq:dx}), the force applied by the actuator was obtained from
the displacement of the test mass multiplied by
16 $ \mu $N/$ \mu $m. 

	After the cryogenic actuator was cooled down to 77 K 
without a magnetic field,
the coil current was raised from 0 to the setting value.
The steps of the measuring procedure were (1) setting 
the coil current, (2) measuring the displacement of the test mass
after the oscillation amplitude of the test mass was reduced sufficiently, 
(3) decreasing the coil current to zero,
and (4) continuing the same procedure for the next coil current.

	When the polarity of the coil current was changed, the direction of 
the force by the actuator to the test mass was not changed.
Also, the values of the forces
were almost the same for both polarities.

	The force by the actuator exceeded 0.1 mN for a coil current of 600 mA
and had 
a maximum value of about 0.2 mN for 1 A.
For coil currents below 400 mA, the measured forces were consistent
with the estimated
values for the ideal superconducting film by Eq.(\ref{eq:calforce}).
	The discrepancies for currents above 400 mA might not be due to 
a decrease of the vortex-free region because, even for a
2A coil current, the applied magnetic field was
about half of the characteristic critical field \cite{Mikheenko,Clem},
which saturated the magnetic
moment of the film with its critical current density; 
moreover, another study showed a larger force \cite{Weinberger}.
	We guess that they
came from a tiny break of the YBCO film by a mismatch of the thermal expansion 
between the YBCO film and the epoxy glue.  	

\section{Discussions}
    \subsection{Requirement for LCGT}

    LCGT may have sapphire substrates for mirrors
with a mass of 54 kg and 1 m long pendulums; thus,
to displace the mirror up to 1 $ \mu $m for fringe lock, the actuator for 
LCGT must apply a force larger than 0.5 mN. With a safety factor, 
a force of 2 mN by the actuator is needed.

    \subsection{Superconducting film for LCGT}

    Unlike conventional actuators with permanent magnets and coils,
the direction of force applied by the superconducting film actuator
to the mirror is only repulsive for a weak applied magnetic field. 
	If we need  an attractive force for fringe lock, we have two examples 
of options.

	One of them is to attach superconducting films to both ends of
the cylindrical substrate of the mirror. A force in the opposite direction can
be applied by the films on the opposite end.
	The experimental results presented  show that the actuator 
with four YBCO films of
1.5 cm$ \times $1.5 cm attached to both ends of the cylindrical 
substrate of the
mirror for LCGT could displace the mirror sufficiently in both
directions.

	Another method is to shift the equilibrium position of the pendulum by
adding a bias current to the coil and to keep the position of the mirror whose
horizontal component of gravitational force is always opposite to
the repulsive force by the actuator. In this method the superconducting films 
may be attached to one end of the cylindrical substrate.
	If YBCO films of the same total area as that of the first example
are attached to one 
side of	the substrate of the LCGT mirror, they would similarly work
as the first example.

	Although the cryogenic actuator which we studied consisted of a thin film,
epoxy glue and the YBCO film, itself, would induce extra thermal noise.
	As for a YBCO film, the critical limit of film thickness deposited 
on the sapphire substrate with a CeO$_2$ buffer layer is 300 nm 
due to a mismatch of the thermal expansion
between the substrate and the YBCO film \cite{THEVA}.
	Thus, we need to further examine whether the $Q$ value does not decrease 
with the YBCO film glued on the test mass and to measure the reduction 
of force 
with a YBCO film of 300 nm thick.

\section{Conclusions}

	We measured the forces applied by a cryogenic actuator with a YBCO film
of 1.6 $ \mu $m in thickness and 0.81 cm$ ^2 $ in area driven 
by a solenoid coil
at 77 K.
The force was applied up to 0.2 mN. This measurement shows that a YBCO film
actuator can be used for fringe lock of the LCGT interferometer
from the viewpoint of force.

\ack
We acknowledge the financial support of the Joint Research and
Development Program of KEK.

% The Appendices part is started with the command \appendix;
% appendix sections are then done as normal sections
% \appendix

% \section{}
% \label{}

\end{document}